\documentclass[aps,pra,floatfix,twocolumn]{revtex4}

\usepackage{graphicx}
\usepackage{epsfig}

\usepackage{url}
\usepackage{paralist}
\usepackage{textcomp}

\begin{document}

\begin{widetext}
\noindent\textbf{Preprint of:}\\
Simon J. Parkin, Timo A. Nieminen, Norman R. Heckenberg,
and Halina Rubinsztein-Dunlop\\
``Optical measurement of torque exerted on an elongated object by
a non-circular laser beam''\\
\textit{Physical Review A} \textbf{70}(2), 023816 (2004)
\end{widetext}


\title{Optical measurement of torque exerted on an elongated object\\ by
a non-circular laser beam}

\author{Simon J. Parkin}
\author{Timo A. Nieminen}
\author{Norman R. Heckenberg}
\author{Halina Rubinsztein-Dunlop}

\date{\today}

\affiliation{Centre for Biophotonics and Laser Science, School of
Physical Sciences, The University of Queensland, QLD 4072,
Australia}


\begin{abstract}
We have developed a scheme to measure the optical torque exerted
by a laser beam on a phase object by measuring the orbital
angular momentum of the transmitted beam. The experiment is a
macroscopic simulation of a situation in optical tweezers, as
orbital angular momentum has been widely used to apply torque to
microscopic objects. A hologram designed to generate LG$_{02}$
modes and a CCD camera are used to detect the orbital component of
the beam. Experimental results agree with theoretical numerical
calculations, and the strength of the orbital component suggest
its usefulness in optical tweezers for micromanipulation.
\end{abstract}

\maketitle

\section{Introduction}
Optical tweezers trap microscopic particles using the gradient
force generated by a tightly focused laser beam \cite{ashkin1986}.
Angular momentum (AM) in the beam can be transferred to the
trapped particle via absorption or scattering. Both spin and
orbital angular momentum have been used to rotate absorbing
particles \cite{he1995,friese1996pra,simpson1997,friese1998ol}.
Spin angular
momentum is due to the polarisation of light, and is $\pm\hbar$
per photon for left or right circularly polarised light
\cite{poynting1909,beth1936}. Angular momentum due to the spatial
distribution of the light's wavefront is called orbital angular
momemtum, and is $l\hbar$ per photon, where $l$ is
the azimuthal mode index \cite{allen1992}.
Polarised light can be used to rotate transparent birefringent
particles~\cite{friese1998nature,higurashi1998b} and
transparent nonspherical particles~\cite{bayoudh2003,bonin2002,bishop2003}.
In both of these cases, the torque is due to the transfer of spin angular
momentum, and can be determined by measuring the degree of
circular polarisation of the light once it has been transmitted
through the particle in the trap~\cite{bishop2003}.

Elongated
particles have also been aligned through the exchange of orbital
angular momentum using non-circular beams
\cite{oneil2002b,santamato2002,dasgupta2003}.
In this case, the gradient forces that
act in optical tweezers to attract a transparent particle towards
regions of high intensity act to rotate the particle so that it
lies within the non-circular high intensity focal spot. The same
effect can be achieved by using two independent beams to trap the ends
of an elongated particle~\cite{bingelyte2003}.
Since this torque arises purely from the interaction between the
particle and the intensity profile of the beam, and is therefore
independent of the polarisation, it depends solely on the transfer of
orbital angular momentum. Notably, when rotating elongated objects,
this torque is much greater than that due to polarisation~\cite{bishop2003},
so the use of orbital angular momentum can be highly desirable.
However, to optically measure the total
angular momentum transferred to the particle, the orbital
component must also be measured. The measurement of this orbital
component is the goal of this present work. However, to avoid the
complication of a highly converging and diverging beam and
microscope optics, a macroscopic experiment is performed rather
than using optical tweezers. This is also desirable to avoid
effects due to spin angular momentum. We simulate the alignment of an
elongated object (a rod) to an elliptical beam on a
macroscopic scale. The torque on the rod can then be determined by
measuring the resulting angular momentum in the beam.

Laguerre--Gauss (LG) modes of laser light with a phase singularity
in the centre of the beam carry orbital AM \cite{allen1992}. These
modes of laser light can be made using computer generated
holograms \cite{he1995b}. A hologram is a recording of the
interference pattern by a light field of interest with a reference
beam. By calculating the interference pattern that results from a
plane wave and LG mode we can make a hologram which will generate
LG modes when illuminated by a Gaussian beam. The same hologram
pattern that was used to make a beam with orbital AM can also be
used to detect orbital AM in a beam as we will demonstrate in this
paper.

Orbital angular momentum states are also of interest to the quantum
information and communication fields as the infinite spatial modes
offer multidimensional entanglement. Computer generated holograms
have been used to generate superpositions of LG modes, and the
same holograms can be used to detect these states. These schemes
have been proposed to measure entanglement on the single photon
level \cite{mair2001,leach2002}.


\section{Theory}

That light and other electromagnetic fields can carry angular
momentum follows directly from the transport of linear momentum,
since the linear and angular momentum flux densities $\mathbf{J}$
and $\mathbf{p}$ are related by
\begin{equation}
\mathbf{J} = \mathbf{r} \times \mathbf{p}. \label{total_am}
\end{equation}
For electromagnetic fields, the momentum flux density is given by
\begin{equation}
\mathbf{p} = \mathbf{S}/c = \mathbf{E} \times \mathbf{H}/c
\end{equation}
where $\mathbf{S}$ is the Poynting vector and $c$ is the speed of
light. The coupled electric and magnetic fields form a spin-1
system, and, in general, (\ref{total_am}) includes both a spin
component, associated with the polarization, and an orbital
component due to the spatial structure of the
field~\cite{humblet1943,crichton2000}.

A monochromatic paraxial field, such as a typical laser beam, can
be specified by a slowly varying complex scalar amplitude function
$u$ that satisfies the scalar paraxial wave equation
~\cite{siegman1986}:
\begin{equation}
\left( \frac{\partial^2}{\partial x^2} +
\frac{\partial^2}{\partial y^2} - 2
\mathrm{i}k\frac{\partial}{\partial z} \right) u = 0. \label{spwe}
\end{equation}
In the paraxial approximation, the two tranverse vector components
of the field de-couple, and the longitudinal component vanishes.
Thus, the two linearly polarized components of the amplitude
individually satisfy the scalar paraxial wave equation, and the
spin and orbital angular momenta de-couple. Henceforth, we will
only consider the orbital angular momentum about the beam axis,
which can be found using the orbital angular momentum operator in
cylindrical coordinates:
\begin{equation}
L_z = -\mathrm{i} \partial/\partial\phi.
\end{equation}
The Laguerre--Gauss modes~\cite{siegman1986} form a complete
orthogonal set of solutions to (\ref{spwe}), so we can write
\begin{equation}
u = \sum_{p=0}^\infty \sum_{l = -\infty}^{l = \infty} a_{pl}
\psi_{pl}
\label{modes1}
\end{equation}
where $a_{pl}$ are mode amplitudes, and
\begin{eqnarray}
\psi_{pl} & = & \left( \frac{2p!}{\pi w^2 (p+|l|)!} \right)^{1/2}
\left( \frac{2r^2}{w^2}\right)^{|l|/2}\nonumber\\ & &
L_p^{|l|}\left( \frac{2r^2}{w^2}\right) \exp\left( -
\frac{2r^2}{w^2}\right) \exp(\mathrm{i}l\phi)
\end{eqnarray}
are the normalized mode functions for Laguerre--Gauss modes
LG$_{pl}$ of degree $p$ and order $l$. Since the LG modes are
orthogonal, the total power is given by
\begin{equation}
P = \sum_{p=0}^\infty \sum_{l = -\infty}^{l = \infty} |a_{pl}|^2.
\end{equation}
Since the mode functions $\psi_{pl}$ are also eigenfunctions of the
orbital angular momentum operator, the orbital angular momentum
flux is simply
\begin{equation}
L = \sum_{p=0}^\infty \sum_{l = -\infty}^{l = \infty} l
|a_{pl}|^2/\omega. \label{oam}
\end{equation}

Notably, the orbital angular momentum depends on the magnitude of
the (complex) mode amplitudes, and not their phase. The power in a particular
mode also depends on the magnitude of the mode amplitude, and hence,
if the modes can be separated, the orbital angular momentum flux
can be determined by measurements of the power. It is not necessary
to separate modes of differing radial degree $p$, only modes of
differing azimuthal order $l$, since the ratio of angular momentum to
power is the same for all modes of the same order $l$. The modes
of differing $l$ can be separated by using a hologram as an
analyzer; this is discussed in the following section, \ref{methodsection}.

The mode amplitudes can also be found directly from the field, if the actual
field is known. In general, this requires knowledge of the actual
electric and magnetic fields, including phase information. However, since
the mode amplitudes themselves are not required, but only their magnitudes,
the holographic filtering we perform provides us with the
necessary information.

In order to theoretically predict the torque exterted on our
test object, we need to know the
angular momentum fluxes of the incident and transmitted beams.
The torque acting on the test object is then given by the
difference in the angular momentum fluxes of the incident and
transmitted beams given by (\ref{oam}).
This requires the magnitudes of the mode amplitudes. We determine the mode
amplitudes of the incident beam by measuring the aspect ratio of our
elliptical incident beam at the beam waist, and assuming that we have
an elliptical Gaussian beam. Thus, by assuming a constant phase across
the beam in the beam waist, we have the required knowledge of the fields.

We determine the mode amplitudes by using an overdetermined point-matching
method, similar to the one we have used previously for
non-paraxial beams~\cite{nieminen2003a}.
Since, for practical computational purposes, the summation in
equation~(\ref{modes1}) must be truncated at a finite degree $p_\mathrm{max}$
and order $l_\mathrm{max}$, we obtain, for a single point
$\mathbf{r}_n = (r_n,\phi_n,z_n)$
\begin{equation}
u(\mathbf{r}_n) = \sum_{p=0}^{p_\mathrm{max}}
\sum_{l = -l_\mathrm{max}}^{l = l_\mathrm{max}} a_{pl}
\psi_{pl}(\mathbf{r}_n).
\label{modes2}
\end{equation}
For a set of $n_\mathrm{max}$ points, this gives a system of linear equations
from which the unknown mode amplitudes $a_{pl}$ can be found. The number
of points $n_\mathrm{max}$ is chosen to be larger than the number of
unknown mode coefficients, which is $(p_\mathrm{max}+1)(2l_\mathrm{max}+1)$,
and $a_{pl}$ are then found numerically using a standard least-squares
solver for an overdetermined linear system.

The use of an overdetermined system eliminates the high-spatial-frequency
artifacts that would otherwise occur if only the minimum possible
number of points was used. The mode amplitudes could also be found using an
integral transform, but the point-matching method allows a coarse grid to
be used and gives good convergence~\cite{nieminen2003a}.

While the incident beam mode amplitudes can be found by measuring
the intensity profile of the incident beam, and assuming a constant phase
in the waist plane, this assumption will not be sufficiently accurate
for the transmitted beam---passage through our test object alters the
phase. Instead, we calculate the complex amplitude (including phase) of
the transmitted beam by treating the
test object as a pure phase object of negligible thickness
altering only the phase of the incident beam as it passes through
(the physical optics approximation).
The same point-matching method is then used to determine the mode amplitudes
of the transmitted beam. Equation (\ref{oam}) then gives the angular
momentum fluxes, and the difference between these is the torque
exerted on the test object.

This technique is used to
calculate the torque as a function of phase thickness
(fig.~\ref{fig:CiTorPh}) and the amplitude of the sinusoidal
variation of torque with respect the angle of the rod in the
elliptical beam (fig.~\ref{fig:sigdif}).

The orbital torque can also be calculated by assuming that the
elongated particle acts as a cylindrical
lens~\cite{santamato2002,beijersbergen1993,courtial1997oc}.
It can be noted that cylindrical lenses can be used as mode
converters to produce Laguerre--Gauss beams which carry orbital
angular momentum, also with resulting orbital
torque~\cite{beijersbergen1993,courtial1997oc}

The elliptical beam is periodic in the azimuthal angle $\phi$,
with period $\pi$. Therefore, a Fourier series expansion of the
azimuthal dependence contains terms of angular frequency $2m$, where
$m$ is an integer. The azimuthal term in the LG modes arises from
such a Fourier expansion, and so, for an elliptical beam, or indeed any
beam with an azimuthal period of $\pi$, $l = 2m$ for non-zero
modes. That is, only modes with even $l$ contribute. The actual
distribution of power among modes of differing $l$ for the incident
beam used in our calculations (a beam with an elliptical focal spot
of aspect ratio 2.25, which is the measured aspect ratio of the beam
used in our experiments), is shown in table \ref{table1}. Almost all of the
power is in modes with $l = 0,\pm 2$.

\begin{table}[htb]
\begin{tabular}{rc}
$l$ & Fraction of power \\
\hline\\
$\ge +4$ & $<1$\% \\
$+2$ & 6.4\% \\
$0$ & 86\% \\
$-2$ & 6.4\% \\
$\le -4$ & $<1$\%
\end{tabular}
\caption{Distribution of power among modes of differing
orbital angular momentum. The beam has an elliptical focal spot, with
an aspect ratio of $2.25$, and has a Gaussian profile along the major
axes of the ellipse.}
\label{table1}
\end{table}

Since the test object we used also has a periodicity in the azimuthal
angle of $\pi$, only even $l$ modes are present in the transmitted
beam. This can be deduced from the fact that, in the physical optics
approximation, the transmitted field is the product of the incident
field and a phase factor $\delta(r,\phi)$:
\begin{equation}
E_\mathrm{trans} = E_\mathrm{inc} \delta.
\end{equation}
Since only the azimuthal variation affects the orbital angular momentum
about the beam axis, it is sufficient to determine the frequencies present
in the Fourier expansion with respect to the azimuthal angle $\phi$.
Since this product is a convolution in the Fourier domain, the
angular frequencies present will be those present in the incident
beam plus those present in the Fourier expansion of $\delta$. Since
both $E_\mathrm{inc}$ and $\delta$ have a period of $\pi$, the convolution
does not alter the angular frequencies present---the sum of two even
integers is still an even integer. If we consider an incident beam that
is a single LG mode, we see that the symmetric scatterer couples the
incident mode to even LG modes in the transmitted beam.

This result does not depend on the physical optics approximation (which
we used above), and is a quite general result relating the rotational
symmetry of a scatterer with the coupling between azimuthal
modes~\cite{schulz1999}.

\section{Method}
\label{methodsection}
We have carried out an experiment designed to measure the orbital
angular momentum component of a light beam, and from this infer
the torque exerted by a light beam on an object in its path. The
orbital angular momentum is detected by a hologram which generates
LG$_{pl}$ modes (say LG$_{0,\pm2}$ in the first order) from a
Gaussian input beam. The LG$_{pl}$ modes have an orbital angular
momentum component of \emph{l}$\hbar$ per photon \cite{allen1992},
so that an LG$_{0,2}$ mode has 2$\hbar$ orbital angular momentum
per photon. If the input beam is instead a LG$_{0,\pm2}$ then a
Gaussian is generated in one of the two first order modes in the
diffraction pattern~\cite{leach2002}.
Therefore if the input into the hologram is
some arbitrary beam, then by measuring the strength of the
Gaussian at the centre of the two first diffraction orders, the
orbital angular momentum carried by the LG$_{0,\pm2}$ components
of the beam can be determined. Only the
Gaussian component in the first diffraction orders has a non-zero
intensity at the centre of the spot. In this experiment the
arbitrary beam is an elliptical beam scattered by a phase object.
The phase object is a bar or rod which is at some angle to the
major axis of the elliptical beam. The orbital angular momentum in
this beam is a result of the various modes which compose the
elliptical beam and the torque exerted on the phase object as it
tends to align with the major axis of the elliptical beam. Due to
the order 2 rotational symmetry of the system, the torque will
predominantly be due to the $l=\pm2$ modes.

\begin{figure}
  \begin{center}
     \includegraphics[width=0.8\columnwidth]{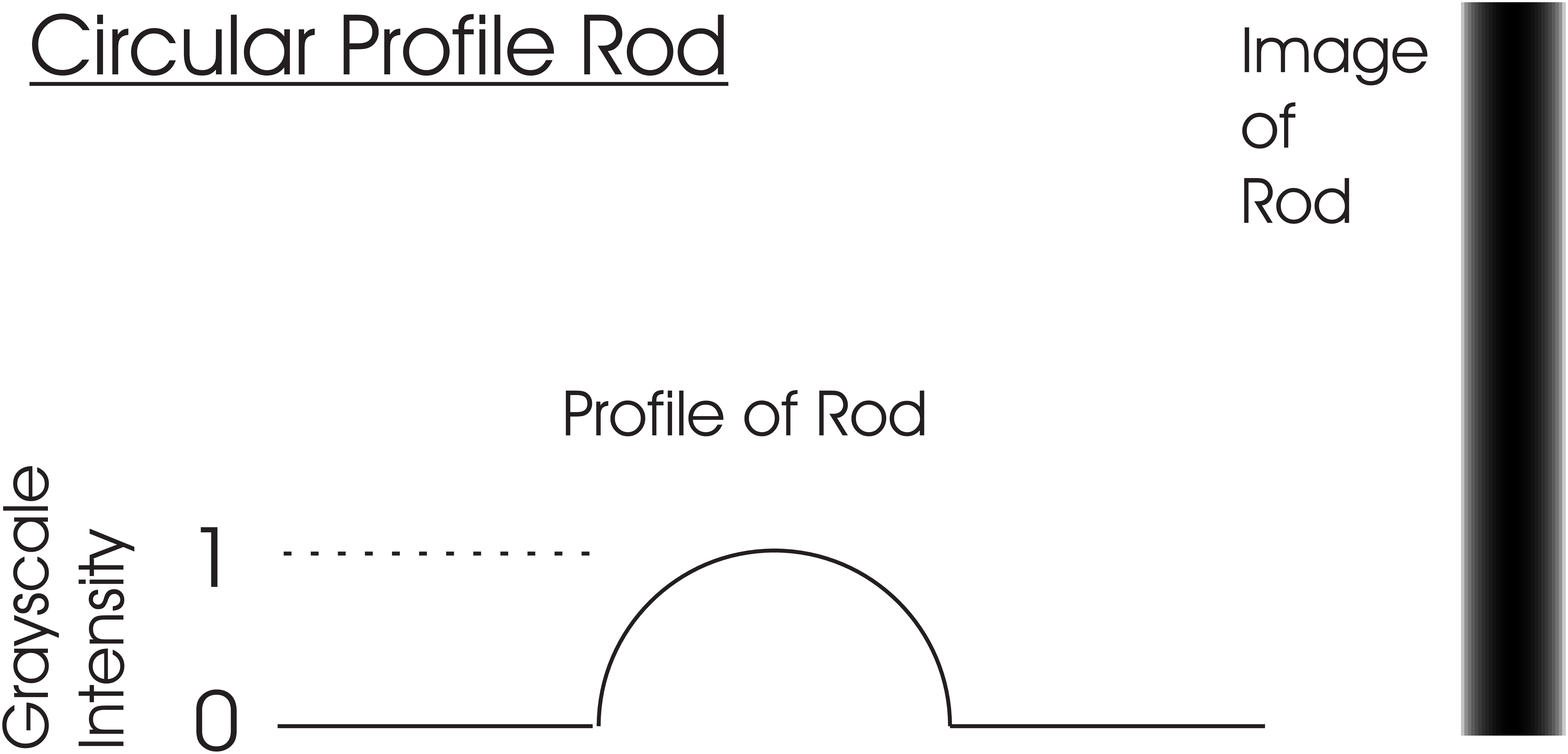}
  \end{center}
  \caption{Grayscale profile and image of the circular rod}
  \label{fig:CnSPro}
\end{figure}

The pattern for the hologram was generated from the calculated
sinusoidal interference resulting from a plane wave and a
LG$_{02}$ mode. This image of the pattern was printed onto film
using a Polaroid ProPalette 7000 Digital Film Recorder. The film
was then contact printed to a holographic plate that has a thick
silver halide emulsion layer. The developed plate was bleached
using mercuric chloride to produce a pattern which acts as a phase
hologram. Images of bars were also made into phase objects using
this same method---the phase picture of the rod was made
from a grayscale image that has a circular profile
(fig.~\ref{fig:CnSPro}).

\begin{figure}[h]
  \begin{center}
   \includegraphics[width=\columnwidth]{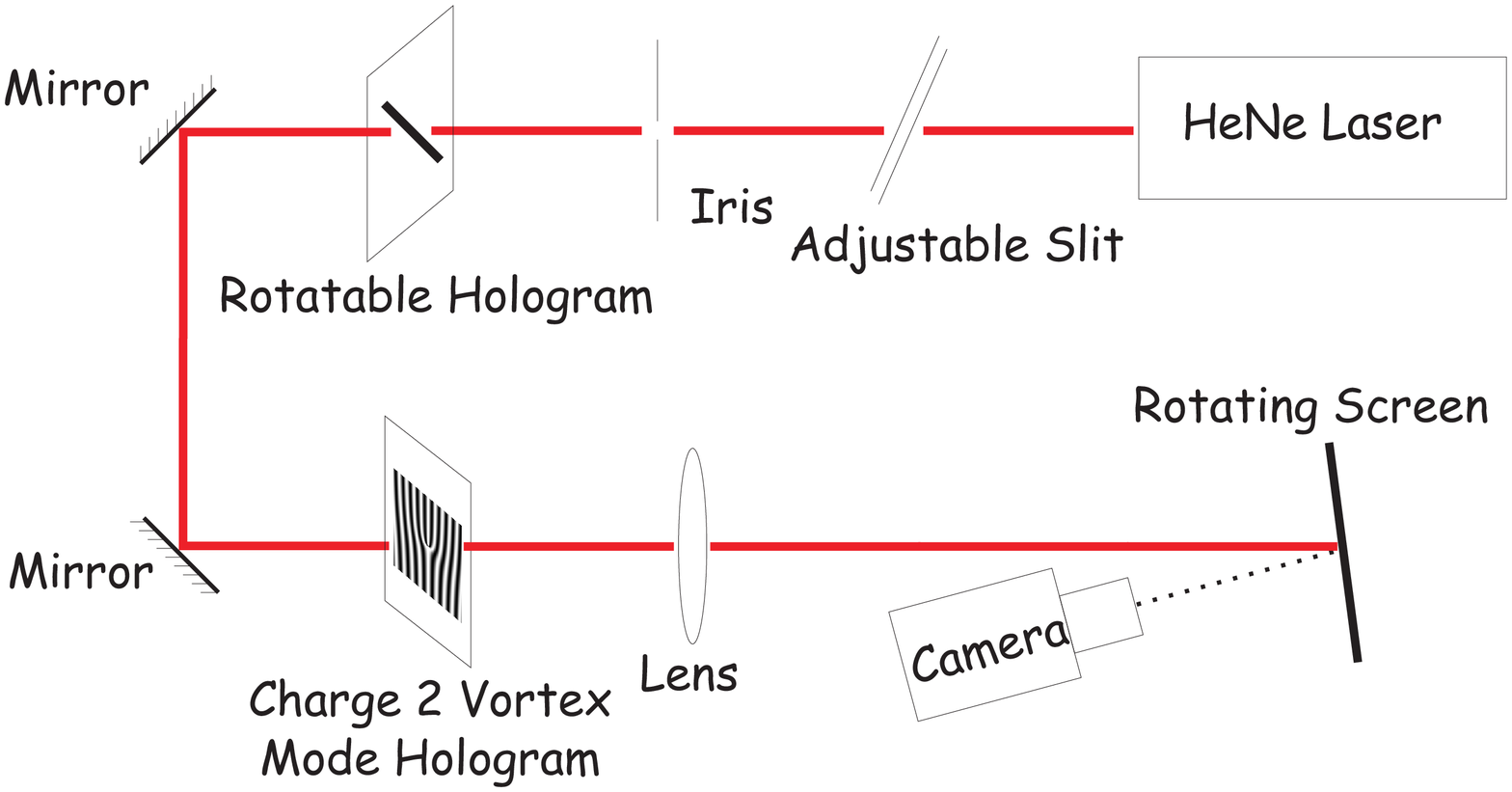}
  \end{center}
  \caption{Experimental setup for measurement of torque on the phase plate (rod)}
  \label{fig:mainsetup}
\end{figure}

The experimental setup is shown in fig.~\ref{fig:mainsetup}. A
helium neon laser beam is directed through an adjustable slit
which creates an elliptical beam that is then incident on a plate.
The plate can be rotated such that the phase image of a rod on the
plate can be oriented at any angle with respect to the beam. The
beam then passes through a second holographic plate which contains
a LG$_{02}$ sinusoidal phase hologram. The beam then passes
through a long focal length lens and onto a rotating screen at the
focal point of the lens. A CCD camera captures the pattern
displayed on the screen. The position of the zero-intensity
spots at the center of each diffraction order is noted when a Gaussian
beam (that is, an LG$_{00}$ beam) is incident on the hologram. The
intensity at these locations is proportional to the power in the
mode with the appropriate angular momentum.
This system is calibrated by measuring the
detected signal produced when the a pure LG mode of known power is
incident on the analysing hologram.

To do this, the slit was removed and another
LG$_{02}$ hologram replaced the phase image of a rod. The first
order mode from the first hologram, which has a known orbital
angular momentum, was then selected and sent through the analysing
hologram. The pattern in the two first order modes from the
analyser was recorded by the CCD camera.

In general, if we consider the measurement of the orbital torque
acting on an arbitrary scatterer rather than an ideal scatterer such
as our phase image, it will not be possible to collect all of the
scattered light. However, likeour phase object,
the transparent particles usually trapped
in optical tweezers do not have a large refractive index contrast
with the surrounding medium, and reflect little of the incident light;
most of the incident light is transmitted through the trapped particle.
Thus, the experiment presented here provides a suitable model for the
measurement of orbital torque in optical tweezers.

\section{Results}

\begin{figure}
  \begin{center}
   \includegraphics[width=0.8\columnwidth]{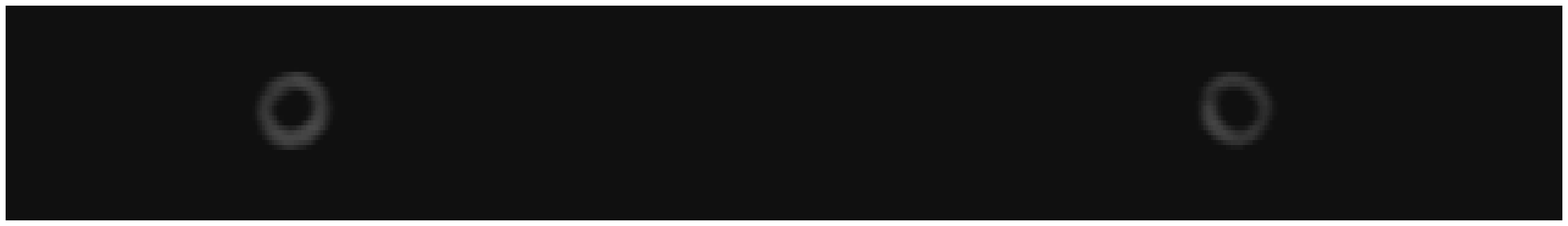}\\
   \includegraphics[angle=270, width=\columnwidth]{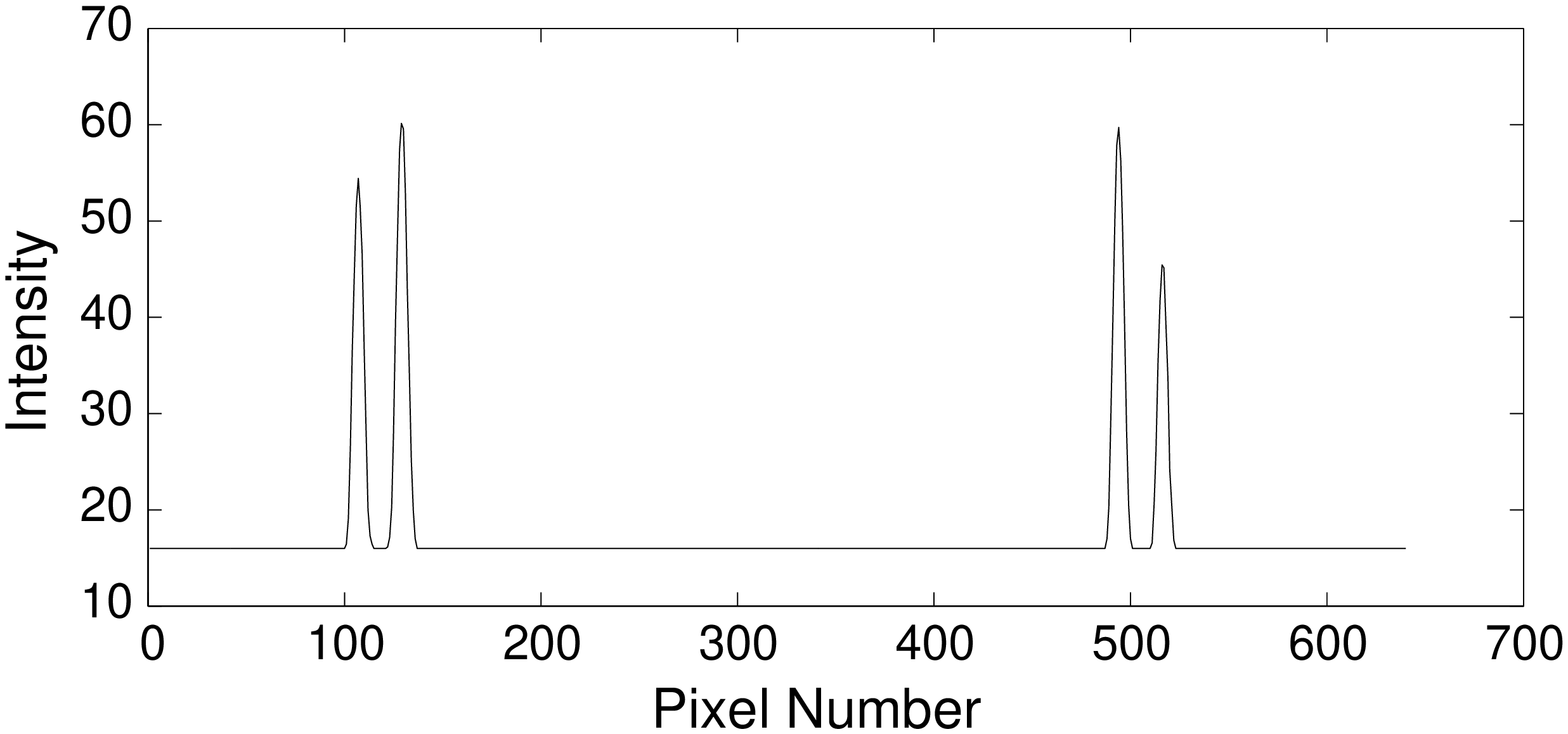}
  \end{center}
  \caption{First order modes from a LG$_{02}$ hologram with a Gaussian input.
  The graph shows a line scan through the image array.}
  \label{fig:Gaussmodepic}
\end{figure}

\begin{figure}
  \begin{center}
   \includegraphics[width=0.8\columnwidth]{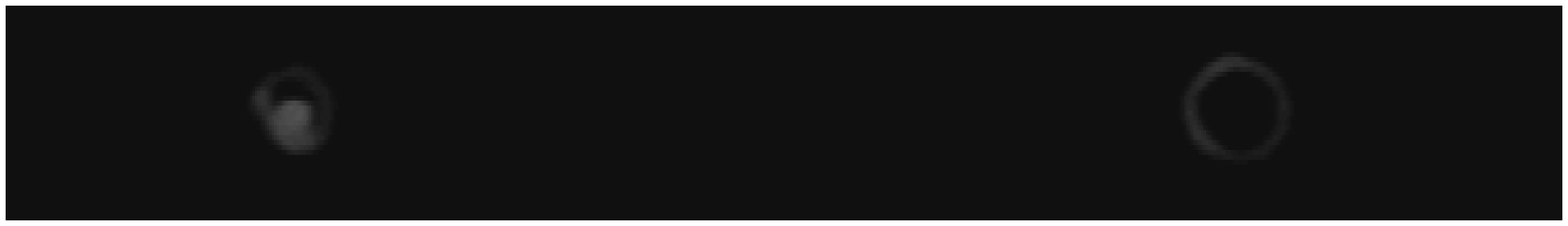}\\
   \includegraphics[angle=270, width=\columnwidth]{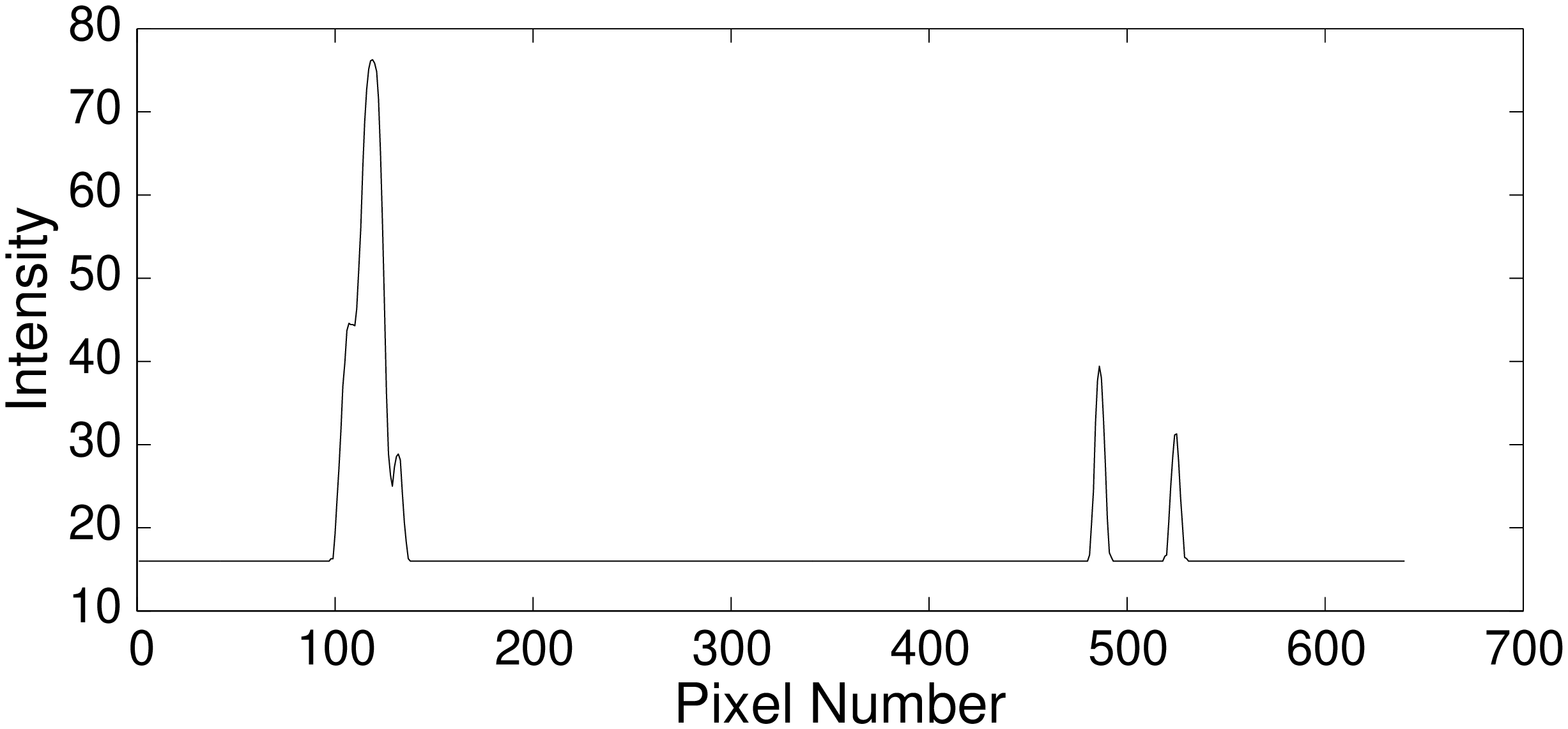}
  \end{center}
  \caption{First order modes from a LG$_{02}$ hologram with a LG$_{02}$ input}
  \label{fig:fillmodepic}
\end{figure}

The two first order modes from the analysing hologram, when the
input is a Gaussian beam, are LG$_{0,+2}$ and LG$_{0,-2}$ modes
(fig.~\ref{fig:Gaussmodepic}). However we see that if an LG$_{02}$
mode is incident on the analyser, one diffracted order from the
analyzing hologram `fills in' to give a
Gaussian and the other is transformed into a higher order LG mode
(fig.~\ref{fig:fillmodepic}). The `filling in' is therefore an
indicator of the angular momentum in the incident beam. With a
Gaussian input, which has no orbital AM, two vortices were
produced at the two first order modes. So the pixels on the CCD
that correspond to the centre of the vortices were then monitored,
as a signal at these centre pixels means that the input beam has
orbital AM. The LG$_{02}$ has a known orbital AM of 2$\hbar$ per
photon, and was used to calibrate the signal at the centre pixels.

\begin{figure}
  \begin{center}
      \includegraphics[angle=270, width=0.48\textwidth]{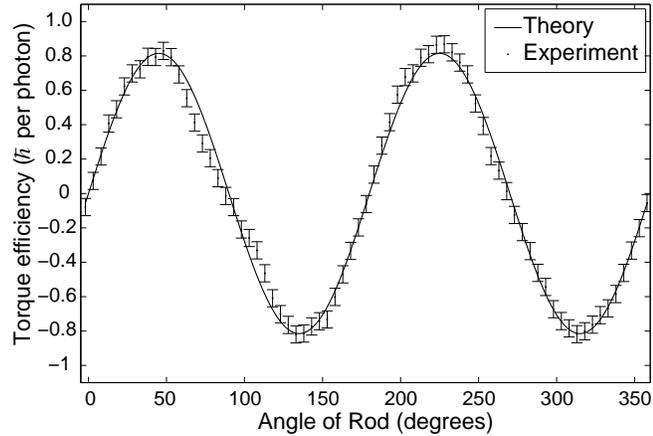}
  \end{center}
  \caption{Signal difference from two centre pixels, for the rotation of a rod
  (with a circular profile) through 360 degrees.}
  \label{fig:sigdif}
\end{figure}

An elliptical beam scattered by a rod at an angle to the beam's
major axis has angular momentum due to the torque tending to align
the rod with the major axis. Monitoring the centre pixels of the
first order modes from the analysing hologram, we were able to
measure the orbital angular momentum flux of the beam, and hence the torque
exerted on the rod. The
difference between the signal at the two centre pixels shows a
sinusoidal variation as the angle of the bar is rotated with
respect to the elliptical beam in agreement with theory
(fig.~\ref{fig:sigdif}). Since the torque is proportional to the beam power,
we show the torque efficiency, given here in units of $\hbar$ per photon.
This is the ratio of the torque to the power divided by the optical
angular frequency ($P/\omega$).

\begin{figure}
  \begin{center}
     \includegraphics[width=\columnwidth]{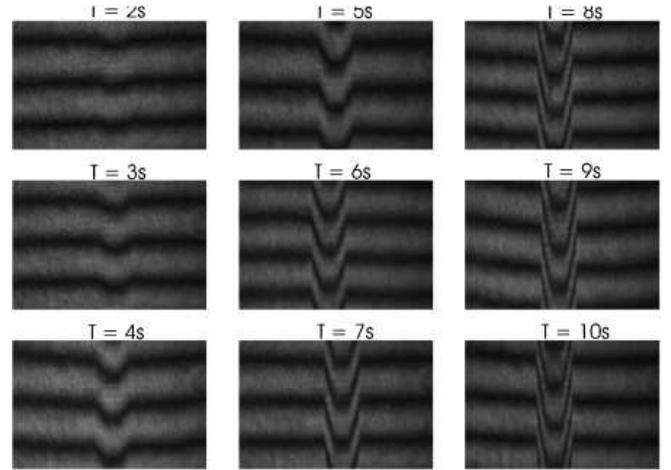}
  \end{center}
  \caption{Interference patterns showing the phase shift of circular profile rods,
  with varying contact print exposure times.}
  \label{fig:Inter9x}
\end{figure}

The torque measured is dependent on the phase thickness of the
rod. The phase thicknesses of a number of rods, that were exposed
for different periods of time during the contact print process,
were measured using a Michelson interferometer.
The phase object was imaged onto a rotating screen and recorded
using a CCD camera. The phase shift of each rod could then be
determined from the shift in fringes of the interference pattern
(fig.~\ref{fig:Inter9x}).

\begin{figure}
  \begin{center}
     \includegraphics[angle=270, width=\columnwidth]{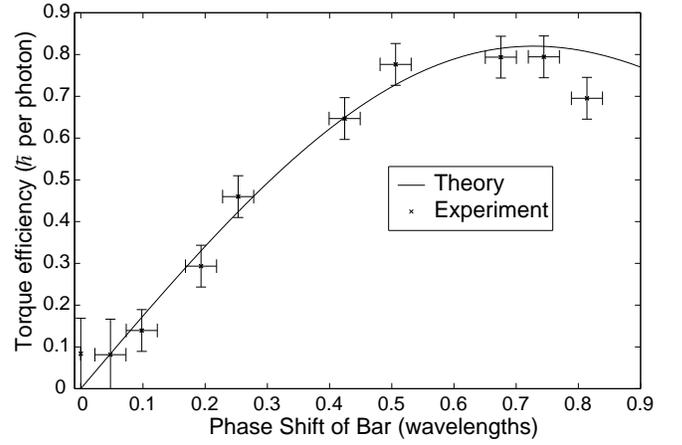}
  \end{center}
  \caption{Signal strength (difference in intensity of the 2 centre pixels)
  relative to phase thickness of different rods (circular profile rods).}
  \label{fig:CiTorPh}
\end{figure}

The rods corresponding to the interference patterns in
fig.~\ref{fig:Inter9x} were placed in the elliptical beam at 45
degrees to the major axis of the elliptical beam when the spatial
torque is greatest. Therefore the torque as a function of phase
shift was found (fig,~\ref{fig:CiTorPh}).

\section{Discussion and Conclusion}

We have shown that in the macroscopic environment, the orbital
angular momentum in a transmitted beam can be measured, allowing
the torque on a phase object to be determined. The theoretical
results show good agreement with the experimental data.

In this experiment the orbital angular momentum transfer
was found to be as
much as 0.8$\hbar$ per photon, compared to 0.05$\hbar$ for the
alignment due to spin angular momentum for a rod in optical
tweezers with a Gaussian beam~\cite{bishop2003}. As the orbital
component is of considerable size it is of potentially useful
technological application if incorporated into optical tweezers.
Also, the effectiveness of this technique to measure orbital
angular momentum allows for complete measurements of the torque in
optical tweezers. So beams that contain an orbital component are
not only useful for micromanipulation, but also the torques
involved can be fully characterised.


\clearpage

\end{document}